\documentclass[twoside]{article}
\usepackage{pmart,fleqn,epsfig}
\begin{document}
\title{EFFECTS OF CORE-HOLE SCREENING ON SPIN-POLARISED AUGER SPECTRA
  FROM FERROMAGNETIC Ni} \author{T. Wegner, M. Potthoff, and W. Nolting}
\address{Lehrstuhl ``Festk{\"o}rpertheorie'', Institut f{\"u}r Physik,
  Humboldt-Universit{\"a}t zu Berlin, Invalidenstra{\ss}e 110, 10115
  Berlin, Germany} \date{April 26, 1999} \maketitle 
\begin{abstract}
  We calculate the spin- and temperature-dependent local density of
  states for ferromagnetic Ni in the presence of a core hole at a
  distinguished site in the lattice. Correlations among the valence
  electrons and between valence and core electrons are described within
  a multi-band Hubbard model which is treated by means of second-order
  perturbation theory around the Hartree-Fock solution. The core-hole
  potential causes strong screening effects in the Ni valence band. The
  local magnetic moment is found to be decreased by a factor 5-6.  The
  consequences for the spin polarisation of CVV Auger electrons are
  discussed.
\end{abstract}
It was pointed out by Allenspach et al.~\cite{AMTL87} that the
experimentally observed spin polarisation of the CVV Auger spectrum of
Ni is substantially smaller than the band spin-polarisation. Contrarily,
in the case of Fe the band and Auger spin-polarisations are comparable.
The authors of ref.~\cite{AMTL87} argue that the ground-state
configuration of Ni in a solid is 3d$^{9}$ which as a consequence of the
core-hole screening becomes a 3d$^{10}$ configuration with a vanishing
magnetic moment. The observed finite Auger polarisation is then
attributed to a core-hole polarisation caused by resonant excitation of
a core electron into the valence band. Furthermore, in the case of Ni
only the minority-spin (3d) states are unoccupied (strong ferromagnet).
While the situation is different in the case of Fe, since there are
unoccupied minority, as well as, majority spin states (weak
ferromagnet).
\\
In the present paper we investigate the effects of core-hole screening
in the initial state of the Auger process quantitatively, i.~e. in a
model of correlated {\em itinerant} electrons. Besides the local
magnetic moment we are interested in the local density of states, which
is relevant for the Auger line shape as is already known from the simple
self-convolution model of Lander~\cite{Lan53}.

We consider a multi-band Hubbard-type model for the 3d, 4s and 4p
electrons. The atomic basis orbitals are assumed to have a well-defined
angular-momentum character $L=\{l,m\}$. The hopping and overlap
integrals for the one-particle part of the Hamiltonian corresponding to
the non-orthogonal atomic basis are taken from (paramagnetic)
tight-binding band-structure calculations~\cite{Pap86}.
\\
The relatively broad 4s and 4p bands are assumed to be sufficiently well
described by band theory.  The Coulomb interaction among the 3d
electrons is assumed to be strongly screened. Consequently, the
interaction part of the Hamiltonian consists of on-site 3d interactions
only.  Exploiting atomic symmetries, the complete Coulomb matrix
$U_{L_1L_2L_4L_3}$ can be expressed via $3j$-symbols~\cite{STK70} in
terms of three effective Slater parameters ($F^0$, $F^2$, $F^4$).
Equivalently, the Coulomb matrix can be parametrised by the averaged
direct and exchange correlation parameters $U$ and $J$ (we take the
atomic ratio for the effective Slater integrals $F^2/F^4$=0.625 which is
a reasonable assumption for 3d-transition metals~\cite{AAL97}).

The one-particle excitation spectrum is calculated by second-order
perturbation theory around the Hartree-Fock solution
(SOPT-HF)~\cite{WPN99}. We furthermore employ the local approximation
since it is known~\cite{SAS92} that the effects due to the (weak) ${\bf
  k}$-dependence of the self-energy are fairly small within SOPT-HF
applied to the multi-band Hubbard model.  The interaction parameters are
chosen as $U$=2.47~eV and $J$=0.5~eV which reproduces the measured value
$m$=0.56~$\mu_{\rm B}$ for the $T$=0 magnetic moment~\cite{HS73}. The
ratio $J/U$$\approx$$0.2$ is a typical value for the late 3d-transition
metals.
\begin{figure}[t]
  \begin{center}
    \epsfig{file=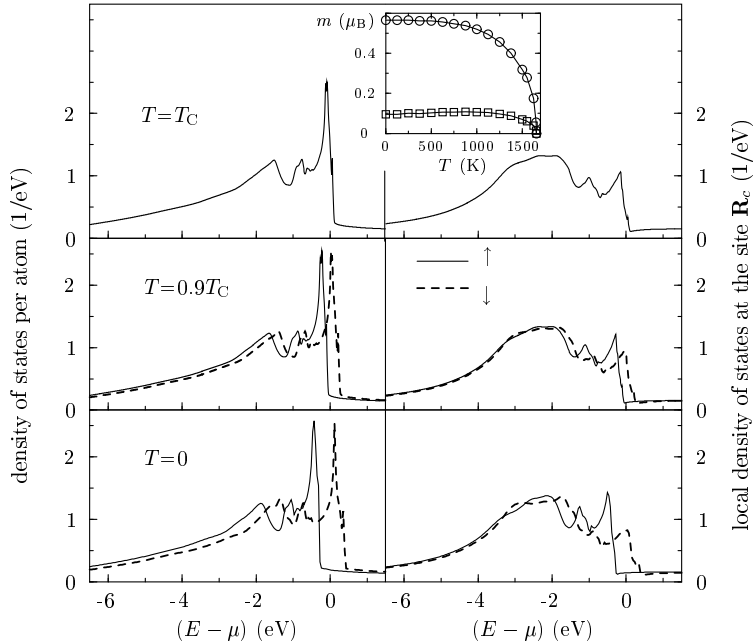, width=0.8\linewidth}
    \caption{Density of states (left) and local density of states at
      the site ${\bf R}_c$ where the core hole was created (right) for
      three temperatures. Solid lines: majority ($\uparrow$) spectrum.
      Dotted lines: minority ($\downarrow$) spectrum. Inset:
      magnetisation (circles) and magnetic moment at ${\bf R}_c$
      (squares) as functions of temperature $T$.}
    \label{figure}
  \end{center}
\end{figure}
\\
The resulting density of states for Ni (fcc) is shown on the l.~h.~s.
of fig.~\ref{figure} for three different temperatures ($T$=0,
$T$=0.9$T_{\rm C}$ and $T$=$T_{\rm C}$).  Due to the imaginary part of
the SOPT-HF self-energy the spectra are strongly damped compared to the
results of band-structure calculations.  In the energy region $\pm$2~eV
around the chemical potential $\mu$ there are distinguishable
structures.  As a consequence of the non-zero slope of the real part of
the self-energy at $E$=$\mu$, a considerable band-narrowing is observed.
The temperature-dependent difference between $\uparrow$ and
$\downarrow$~spectra is more or less given by a rigid shift which
disappears for $T$$\rightarrow$$T_{\rm C}$. The corresponding
magnetisation curve (circles in the inset of fig.~\ref{figure}) has a
Brillouin-function-like form.  The Curie temperature turns out to be
$T_{\rm C}$=1655~K and is thereby about a factor 2.6 larger than the
measured value of 624~K~\cite{LB19a}.  This overestimation of $T_{\rm
  C}$ is probably due to the mean-field character of the SOPT-HF.  It
should be noted that a simple LDA+$U$ (Hartree-Fock) calculation yields
a much higher value ($T_{\rm C}$$\approx$2500~K).

Conceptually, the Auger process can be divided into two subprocesses.
The first one is the creation of a core hole at a particular lattice
site ${\bf R}_c$, e.~g. by absorbing an x-ray quantum.  The second one
is the radiationless decay of the core hole by ejecting an Auger
electron.  Provided that the life time of the core hole is large
compared to typical relaxation times of the valence electrons, the two
subprocesses become independent~\cite{AH83}.  Since the Auger process
takes place locally, the Auger spectrum is influenced by the additional
core-hole potential in the initial state.

To describe the core-hole effects we have to extend the Hamiltonian. In
the one-particle part we additionally consider a non-degenerate (s-like)
and dispersionless core level with a one-particle energy well below the
valence band. In the interaction part we add a density-density
interaction between core and valence electrons. This interaction is
responsible for the screening of the core-hole potential.  The
corresponding Coulomb-matrix elements are taken to be orbitally
independent ($U_L^c$=$U_c$). We assume an infinite life-time of the core
hole in the initial state for the Auger process as there are no decay
terms in the Hamiltonian. Thus, the core-level occupation is a good
quantum number.  Thermodynamic averages in the presence of the core hole
have to be performed in the subspace of the Hilbert space that is built
up by all many-body states with a core hole at the site ${\bf R}_c$.  In
practice, this is done by introducing an appropriate Lagrange parameter.

The extended Hamiltonian and the averaging procedure introduces two new
terms in the valence-band self-energy.  The Hartree-like term
$-\delta_{{\bf R R}_c}$$U_c$ represents the additional core-hole
potential seen by the valence electrons.  Thereby, the translational
symmetry is broken and all occupation numbers become site dependent.
Since the correlation effects among the valence electrons depend on the
occupations, this introduces an extra screening term in the self-energy.
\\
These screening effects are in general extended over some shells around
${\bf R}_c$. A reasonable approximation for 3d-transition metals is to
assume a complete screening of the core-hole potential already at the
site ${\bf R}_c$, i.~e. the total occupation at ${\bf R}_c$ is increased
by 1 electron, and one is left with a single-site scattering problem
(for a more detailed discussion see ref.~\cite{WPN99}).  Here, the
complete screening is used as a condition to fix the core-valence
interaction parameter $U_c$ at $T$=0. We find the value $U_c$=1.81~eV.
\\
The local density of states at the site ${\bf R}_c$ in the presence of
the core hole is shown on the r.~h.~s. of fig.~\ref{figure}. The
structure of the spectrum has remarkably changed. Spectral weight is
transferred to lower energies. Especially for $\downarrow$ electrons a
redistribution from energies above to below the chemical potential $\mu$
is visible. By comparing the quasi-particle weights (band-width
renormalisation) with the unscreened case, we find the screened case to
behave less correlated since here one is closer to the limit of the
completely filled (3d) band. As a consequence of the fact that Ni is a
strong ferromagnet, only minority spin states can be populated to screen
the core hole. Indeed this leads to a drastic reduction of the local
magnetic moment (0.095~$\mu_{\rm B}$).  The temperature dependence of
the local magnetic moment in the presence of the core hole is shown in
the inset (squares).

We conclude that the comparatively small spin polarisation of Auger
electrons for Ni~\cite{AMTL87} is due to the screening of the core-hole
potential in the initial state. Within the considered itinerant-electron
model including 4s and 4p states, however, even a complete screening
does not lead to fully vanishing local magnetic moment.

{\bf Acknowledgements:} Financial support of the \textit{Deutsche
  Forschungsgemeinschaft} within the project No.~158/5-1 is greatfully
acknowledged.  The numerical calculations were performed on a CrayT3E at
the \textit{Konrad-Zuse-Zentrum f\"ur Informationstechnologie Berlin
  (ZIB)}.


\begin{thebibliography}{10}
  
\bibitem{AMTL87} R. Allenspach, D. Mauri, M. Taborelli, and M. Landolt,
  Phys. Rev. B {\bf 35}, 4801 (1987).
  
\bibitem{Lan53} J.~J. Lander, Phys. Rev. {\bf 91}, 1382 (1953).
  
\bibitem{Pap86} D.~A. Papaconstantopoulos, {\em Handbook of the band
    structure of elemental solids} (Plenum, New York, 1986).
  
\bibitem{STK70} S. Sugano, Y. Tanabe, and H. Kamimura, {\em Multiplets
    of transition-metal ions in crystals}, Vol.~33 of {\em Pure and
    applied physics} (Academic, New York, 1970).
  
\bibitem{AAL97} V.~I. Anisimov, F. Aryasetiawan, and A.~I. Lichtenstein,
  J. Phys.: Condens.  Matter {\bf 9}, 767 (1997).
  
\bibitem{WPN99} T. Wegner, M. Potthoff, and W. Nolting, (to be
  published).
  
\bibitem{SAS92} M.~M. Steiner, R.~C. Albers, and L.~J. Sham, Phys. Rev.
  B {\bf 45}, 13272 (1992).
  
\bibitem{HS73} S. Hirooka and M. Shimizu, Phys. Lett. {\bf 46A}, 209
  (1973).
  
\bibitem{LB19a} M.~B. Stearns, in {\em Landoldt-B{\"o}rnstein, New
    Series}, Vol.~19a of {\em Group III}, edited by H.~P.~J. Wijn
  (Springer, Berlin, 1984), Chap.~Magnetic Properties of Metals.
  
\bibitem{AH83} C.-O. Almbladh and L. Hedin, in {\em Handbook on
    Synchrotron Radiation}, edited by E.-E. Koch (North-Holland,
  Amsterdam, 1983), Vol.~1b, p.\ 607.

\end{thebibliography}
\end{document}